\documentclass[twocolumn]{aa}    
\usepackage{graphicx,rotating}
\usepackage{subfig}
\usepackage{txfonts}
\usepackage[T1]{fontenc}
\usepackage{epsfig}
\usepackage{natbib}
\usepackage{color}
\usepackage{natbib}
\usepackage{amssymb}  
\usepackage{tabularx}
\DeclareGraphicsExtensions{.pdf,.png,.jpg}
\usepackage[nomarkers,nolists]{endfloat}
\bibpunct{(}{)}{;}{a}{}{,} 
\def \fermi {{\it Fermi}}
\def \pks  {PKS~1510--089}
\def \integral {{\em INTEGRAL}}

\begin{document}

\title{Multiwavelength variability study and search for periodicity of PKS~1510~-089}

\author{G. Castignani\inst{1,2,3,4}
\and
E. Pian\inst{5,6,7}
\and
T.~M. Belloni\inst{8}
\and
F. D'Ammando\inst{9,10}
\and
L. Foschini\inst{8}
\and
G. Ghisellini \inst{8}
\and
T. Pursimo\inst{11}
\and
A. Bazzano\inst{12}
\and
V. Beckmann\inst{13}
\and
V. Bianchin\inst{5}	
\and
M.~T. Fiocchi\inst{12}
\and
D. Impiombato\inst{14}
\and
C.~M.  Raiteri\inst{15}
\and
S. Soldi\inst{16}
\and
G. Tagliaferri\inst{8}
\and 
A. Treves\inst{17}
\and
M. T\"urler\inst{18}
}

\institute{
    LERMA, Observatoire de Paris, CNRS, 61 avenue de l'Observatoire, 75014 Paris, France\\
    \email{gianluca.castignani@obspm.fr}    
    \and
    Coll\`{e}ge de France, 11 place Marcelin Berthelot, 75005 Paris, France
    \and
    Laboratoire Lagrange, Universit\'e C\^ote d'Azur, Observatoire de la C\^ote d'Azur, CNRS, 
Blvd de l'Observatoire, CS 34229, 06304 Nice cedex 4, France
    \and	
    Centre National d'\'{E}tudes Spatiales, CNES, 2 Place Maurice Quentin, 75001 Paris, France
    \and
    INAF - Istituto di Astrofisica Spaziale e Fisica Cosmica, Sezione di Bologna, via Gobetti 101, 40129 Bologna, Italy
      \and
    Scuola Normale Superiore, Piazza dei Cavalieri 7, 56126 Pisa, Italy
    \and
    INFN - Sezione di Pisa, Largo B. Pontecorvo 3, 56127 Pisa, Italy
    \and
    INAF,  Osservatorio Astronomico di Brera, Via E. Bianchi 46, 23807 Merate (LC), Italy
    \and
    INAF-IRA, Via P. Gobetti 101, 40129 Bologna, Italy
    \and
    Dipartimento di Fisica e Astronomia, Universit\`{a} degli Studi di Bologna, Viale Berti Pichat 6/2, Bologna 40127, Italy
    \and                                                                         
    Nordic Optical Telescope, Apartado 474, 38700, Santa Cruz de La Palma, Spain
    \and
    INAF - Istituto di Astrofisica e Planetologia Spaziali, via Fosso del Cavaliere 100, 00133, Roma, Italy
   \and
    CNRS / IN2P3, 3 rue Michel Ange, 75794 Paris Cedex 16, France
     \and
    INAF-IASF Palermo, Via Ugo La Malfa 143, Palermo, I-90146, Italy
    \and
    INAF,  Osservatorio Astrofisico di Torino,  Via Osservatorio 20, 10025 Pino Torinese (TO),  Italy
    \and
   Centre Fran\c{c}ois Arago, APC, Universit\'e Paris Diderot, CNRS/IN2P3, 10 rue Alice Domon et
    L\'eonie Duquet, 75205 Paris Cedex 13, France
    \and
    Universit\`a degli Studi dell'Insubria, via Valleggio 11, 22100, Como, Italy
    \and
    {\it INTEGRAL} Science Data Centre, University of Geneva, Chemin d'Ecogia 16, 1290 Versoix, Switzerland
    }

\date{}	

\abstract
   {Blazars are the most luminous and variable active galactic nuclei (AGNs), and thus excellent probes of 
   accretion and emission processes close to the central engine.}
   {We concentrate here on \pks\ ($z = 0.36$), a member of the Flat-Spectrum Radio Quasar variety of blazars, an extremely  powerful gamma-ray source and one of the brightest in the  \fermi\ LAT catalog, to study the complex variability of its bright multi-wavelength spectrum, identify the physical parameters responsible for the variations and the time scales of possible recurrence and quasi-periodicity at high energies.}
{\pks\ was observed twice in hard X-rays with the IBIS instrument onboard INTEGRAL during the flares of
Jan 2009 and Jan 2010, and simultaneously with Swift and the Nordic Optical Telescope (NOT), in addition to the constant Fermi monitoring. We also measured the optical polarization in several bands on 18 Jan 2010 at the NOT.   { Using these and archival data we constructed historical light curves at gamma-to-radio wavelengths covering nearly 20 years and applied  tests of fractional and correlated variability. We also assembled spectral energy distributions (SEDs) based on these data and compared them with those at two previous epochs, by applying a model based on synchrotron and inverse Compton radiation from blazars.}}
   {The modeling of the SEDs suggests that the physical quantities  that  undergo the largest variations are the total power injected into the emitting region and the random Lorentz factor of the electron distribution cooling break, that are  higher in the higher gamma-ray states. This suggests a correlation of the injected power with enhanced activity of the acceleration mechanism.   The cooling likely takes place at a distance of  $\sim$1000 {Schwarzschild} radii { ($\sim$0.03~pc)} from the central engine, i.e much { smaller} than the broad line region (BLR) radius. 
{ The  emission at a few hundred GeV can be reproduced with inverse Compton scattering of highly relativistic electrons off far-infrared photons if these are located much farther than the BLR, i.e.,  around 0.2~pc from the AGN, presumably in a dusty torus.}
We determine a luminosity of the thermal component due to the inner accretion disk of 
$L_d  \simeq  5.9\times10^{45}$~erg~s$^{-1}$, a BLR luminosity of $L_{\rm BLR} \simeq 5.3 \times 10^{44}$~erg~s$^{-1}$, and a mass of the central black hole of $M_{\rm BH} \simeq 3\times10^{8}$~M$_\odot$. 
{ The fractional variability as a function of wavelength follows the trend expected if X- and gamma-rays are produced by the same electrons as radio and optical photons, respectively.}
{ Discrete Correlation Function (DCF) analysis between the long-term Steward observatory optical V-band and gamma-ray \fermi\ LAT light curves yields a good correlation with no measurable delay. Marginal correlation where X-ray photons lag both optical and gamma-ray ones by time lags between 50 and 300~days is found with the DCF. }
Our time analysis of the RXTE PCA  and  \fermi\ LAT light curves reveals no  obvious (quasi-)periodicities, at least up to the maximum time scale (a few years) probed by the light curves, which are  severely affected by red noise.}   
   {}


\authorrunning{Castignani et al.}
\titlerunning{Multiwavelength variability study and search for periodicity of PKS~1510~-089}

\maketitle

\section{Introduction}
Blazars are among the  most luminous  persistent extragalactic multi-wavelength sources in the Universe.
{Their gamma-ray radiation is likely produced within hundreds to a few thousands of Schwarzschild radii of the central engine \citep{Finke_Dermer2010,Ghisellini2010,Tavecchio2010} or at the location of the circum-nuclear dust at parsec scales \citep{Marscher2008,Sikora2009,Marscher2010}.}
Therefore blazar variability  on a range of time-scales from years down to hours is {one of the most powerful} diagnostic of their emission geometry and mechanisms  \citep[see e.g.,][]{Boettcher_Chiang2002,Boettcher_Dermer2010,Hayashida2012,Falomo2014,Dermer2014,Chidiac2016,Krauss2016}.

\pks\ is a nearby ($z=0.36$), variable, highly polarized   Flat-Spectrum Radio Quasar  \citep[FSRQ,][]{Burbidge_Kinman1966,Stockman1984,Hewitt1993}, 
well monitored  at all bands from the radio to gamma rays \citep[e.g.,][]{Malkan1986,Thompson1993,Pian1993,Sambruna1994,Lawson1997,Singh1997,Siebert1998,Tavecchio2000,Dai2001,Gambill2003,Wu2005,Bach2007,Li2007,Kataoka2008,Nieppola2008}.  
It is  one of the brightest and most variable blazars  detected by \fermi\ LAT  \citep{Abdo2009,Abdo2010a,Abdo2010e,Foschini2013,Ackermann2015a} and AGILE-GRID \citep{Pucella2008,Dammando2009,Dammando2011}, and one of the only { six} FSRQs detected at very high energies, i.e., up to a few hundred GeV in the observer frame, with 3C~279, 4C~21.35, PKS~1441~+25, { S3~0218+35, and PKS~0736+017} \citep[][]{Cohen2003,Albert2008,Aleksic2011,Abramowski2013,Abeysekara2015,Ahnen2015,Cerruti2016}.

For these reasons, it was the target of many recent multi-wavelength observing campaigns, some of which included monitoring of its highly variable optical polarization \citep[e.g.,][]{Rani2010,Abdo2010b,Abdo2010c,Marscher2010,Ghisellini2010,Sasada2011,Orienti2011,Arshakian2012,Orienti2013,Stroh_Falcone2013,Aleksic2014,Fuhrmann2016}.  

This rich dataset became  a benchmark for phenomenological studies of \pks\  \citep[e.g.,][]{Kushwaha2016},    theoretical modeling of its  light curves \citep{Tavecchio2010,Brown2013,Cabrera2013,Nakagawa_Mori2013,Nalewajko2013,Saito2013,Vovk_Neronov2013,Marscher2014,Dotson2015,MacDonald2015,Kohler_Nalewajko2015},  and  spectral energy distributions \citep[SEDs,][]{Abdo2010c,Chen2012,Nalewajko2012,Yan2012,Boettcher2013,Barnacka2014,Nalewajko2014, Saito2015,Boettcher_Els2016,Basumallick2016}.


Evidence for possible year time-scale periodicity of \pks\  was  found { from the}  radio to gamma rays \citep[e.g.,][]{Xie2008,Abdo2010c,Castignani2010,Sandrinelli2016}.   The  search for  quasi-periodicities  in light  curves  \citep[e.g.,][]{Valtonen2008,Valtonen2011,Zhang2014,Sandrinelli2014a,Sandrinelli2014b,Graham2015a,Graham2015b,Ackermann2015b} and spectra  \citep[e.g.,][]{Tsalmantza2011,Ju2013,Shen2013} of AGNs
is a long-standing issue because it is related to the possible presence of  close binary systems of
supermassive black holes (SMBHs) 
and  received recent  boost from the relevance of  these systems as potential sources of  milli-Hz gravitational  waves \citep{Amaro-Seoane2013,Colpi2014}.

Because of its characteristics \pks\ is a prominent target for our  program of   \integral\   observations of blazars in outburst.  The source was first observed by \integral\ IBIS in 2008 \citep{Barnacka2009} and detected with a hard X-ray flux that was about a factor of two lower than observed by {\it Suzaku} in 2006 \citep{Kataoka2008}.  We have re-observed it during high states in Jan 2009 and Jan 2010, and
report here a study of  its multi-wavelength light curves and spectra at various epochs, based on the \integral\ data and quasi-simultaneous observations made with the
{\it Swift} satellite  (some of which are presented here for the first time), {\it Fermi}-LAT and with the 
Nordic Optical Telescope (NOT).  
In order to put our data in the context of the long-term behavior of \pks\ we have also retrieved from the archive and the literature { gamma-ray (EGRET,  \fermi,  and AGILE), optical (Steward Observatory and the Rapid Eye Mount telescope), and radio (MOJAVE VLBA program) data of this source.} 
Moreover,   we have reanalyzed the  RXTE PCA dataset of \pks\  up to Apr 2011 and  present a timing analysis of this light curve in search of periodicity  using several independent methods.
We also present models of the SEDs that we have elaborated to identify the main parameters that are responsible for the multi-wavelength long-term variability.

Throughout this work we adopt a flat $\Lambda \rm CDM$ cosmology with matter density $\Omega_{\rm m} = 0.30$, dark energy density $\Omega_{\Lambda} = 0.70$ and Hubble constant $h=H_0/100\, \rm km\,s^{-1}\,Mpc^{-1} = 0.70$ \citep[see however,][]{PlanckCollaborationXIII2015,Riess2016}.  Under these assumptions the luminosity distance of \pks\ is 2.0~Gpc.

In Sect.~\ref{sec:data} we report  the observations of our campaigns, the data analysis, and the results. 
In Sect.~\ref{sec:multilambda_variability} we describe the multiwavelength variability and cross-correlation among light curves, and  present the  results of our periodicity search.
Sect.~\ref{sec:SEDs}  reports on  multi-epoch SED  construction and  modeling. In Sect.~\ref{sec:conclusions} we draw our conclusions.

\section{Observations, data analysis and results}\label{sec:data}

Our \integral\ program for blazars in outburst as Targets of Opportunity was activated in 2009 following a notification of a rapid GeV flare of \pks\ detected by the LAT instrument on board the \fermi\ gamma-ray satellite on Jan 8, 2009 \citep{Ciprini_Corbel2009,Abdo2010c}, and again in 2010 after the detection of an outburst at  $E>100$~MeV with the GRID instrument on-board AGILE  on Jan 11-13, 2010  \citep{Striani2010}.
Multi-wavelength observations with the {\it Swift} satellite  and the NOT were coordinated quasi-simultaneously with the \integral\ campaign.
RXTE PCA  and {\it Fermi} LAT data retrieved from the archive were also used to study the variability over an extended time interval. 
In the following we describe the data reduction and analysis of the multi-wavelength data.  

\begin{figure*} \centering
\subfloat{\includegraphics[width=0.5\textwidth,natwidth=610,natheight=642]{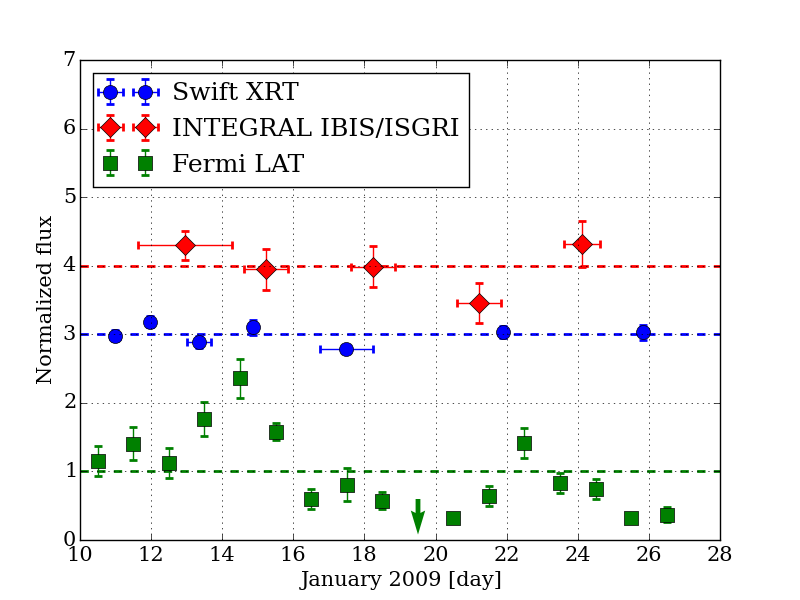}}
\subfloat{\includegraphics[width=0.5\textwidth,natwidth=610,natheight=642]{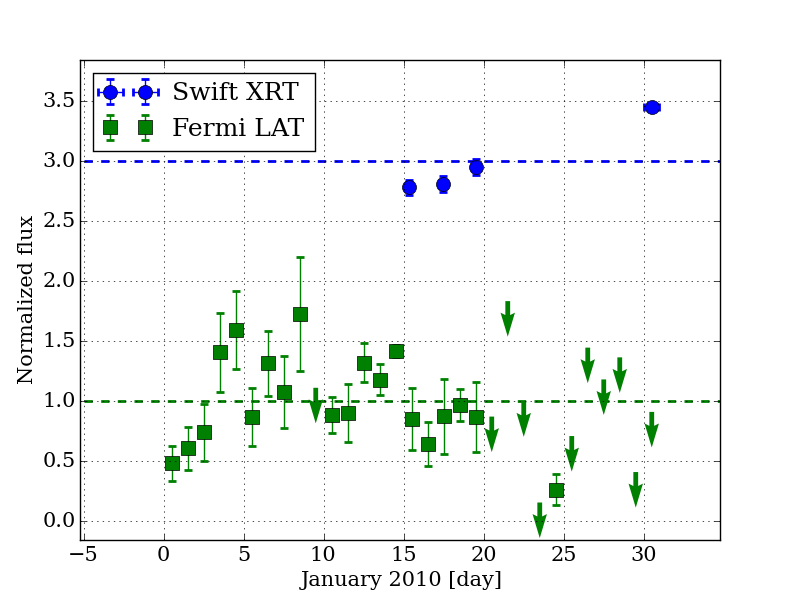}}
\caption{Normalized light curves in observer frame during Jan 2009 (left) and Jan 2010 (right).
{\it Swift} XRT (0.3-10 keV) data, corrected for Galactic HI absorption,
are filled circles (blue); diamonds (red) are \integral\ IBIS/ISGRI
(20-100 keV) data; squares (green) represent the \fermi\ LAT (0.1-300
GeV) data
daily-binned light curves, including 90\% upper limits. All light curves
are normalized with respect to their mean (computed without considering
upper limits). 
{ For each light curve the dashed horizontal line shows the average normalized flux.}
{\it Swift} XRT and \integral\ normalized light curves
are shifted up by constants 2 and 3, respectively.}
\label{fig:lcurve_jan2009_jan2010}
\end{figure*}

\begin{figure*}[htbp]  
\begin{center}
\includegraphics[width=1.0\textwidth]{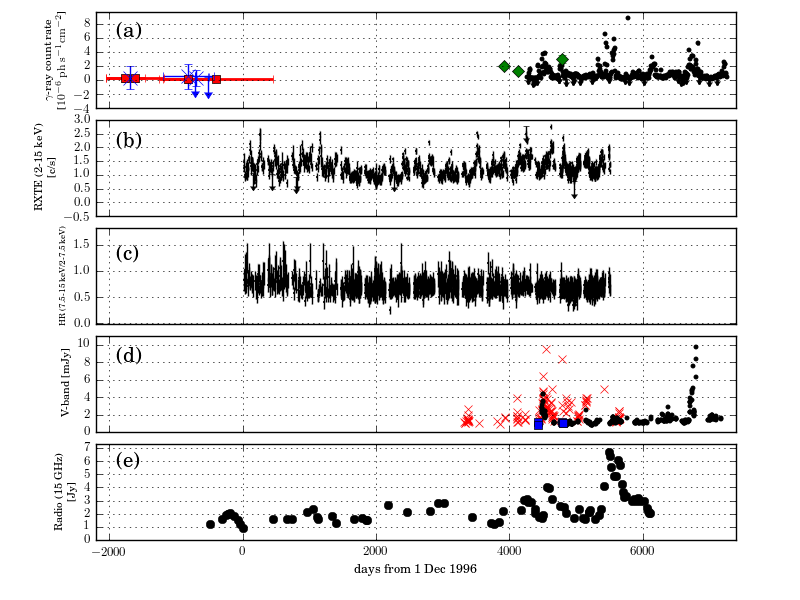}
\caption{Historical light curves  in observer frame.
(a): gamma-ray light curve. The small filled circles (black) represent the {\it Fermi}-LAT 
(0.1-300~GeV) weekly-binned light curve; the 90\% confidence upper limits are also reported;
crosses and big upper limits (blue) represent the CGRO-EGRET ($>$100~MeV) fluxes from 3rd~EGRET catalog \citep{Hartman1999};  squares (red) are the CGRO-EGRET fluxes from the revised EGRET catalog \citep{Casandjian08};  diamonds (green) are AGILE-GRID ($>$100~MeV)  detections \citep{Pucella2008,Dammando2009,Striani2010}. 
(b): X-ray Rossi XTE light curve (2-15 keV) and 3-$\sigma$ upper limits. { The average (median) separation between consecutive observations, once upper limits are removed, is $4.2\pm6.5$~days (3.4~days). The reported uncertainty is the rms dispersion.}
(c): X-ray  Rossi XTE hardness ratio { (7.5-15 keV vs 2-7.5 keV channels)}.
(d): optical V-band historical light curve, corrected for Galactic dust absorption \citep[$E_{B-V} = 0.09$,][]{Schlafly_Finkbeiner2011}, using the extinction curve from \citet{Cardelli1989}. The REM observations (red crosses) are from \citep[][]{Sandrinelli2014a}; the black circles represent photometry obtained at the Steward Observatory \citep{Smith2009}; the blue squares are our NOT data (Tab.~\ref{table:NOTphotometry}).  
(e): radio VLBA I-band (15~GHz) light curve from the MOJAVE database \citep{Lister2009}.
}
\label{fig:hist.lcurve}
\end{center}
\end{figure*}

In Fig.~\ref{fig:lcurve_jan2009_jan2010} we report the high-energy light curves of the 2009 and 2010 campaigns and in Fig.~\ref{fig:hist.lcurve} the historical multi-wavelength light curves.
{ The \integral\ IBIS/ISGRI data of Jan 2010 have sufficient signal-to-noise ratio (SNR) to produce an integrated spectrum, but not a light curve. Therefore they are not plotted in the right panel of Fig.~\ref{fig:lcurve_jan2009_jan2010}.}
In the following sub-sections we will present the gamma-ray-to-optical data. The radio light curve (Fig.~\ref{fig:hist.lcurve}e) is from VLBA measurements in I-band (15 GHz) downloaded from the MOJAVE database\footnote{http://www.physics.purdue.edu/astro/MOJAVE/sourcepages/1510-089.shtml}. 

\subsection{Gamma-rays}\label{par:gamma_rays}            

Gamma-ray data of \pks\ were downloaded from the \fermi\ LAT public archive
\footnote{http://fermi.gsfc.nasa.gov/ssc/data/access/lat/msl\_lc/}
in the form of weekly- and daily-binned light curves.  
In panel~(a) of Fig.~\ref{fig:hist.lcurve} we show the 0.1-300~GeV weekly-binned $\sim8$~years-long \fermi\ LAT light curve, where flux measurements are reported along with 90$\%$ upper limits.
EGRET ($>$100~MeV) data \citep{Hartman1999,Casandjian08} and AGILE-GRID observations of summer 2007 \citep{Pucella2008} and Mar 2008 \citep{Dammando2009} are also reported.
The daily-binned \fermi\ LAT 0.1-300~GeV light curves during the flaring periods of 2009 Jan 11 - Feb 1 and 2009 Dec 26 - 2010 Jan 23 are reported in Fig.~\ref{fig:lcurve_jan2009_jan2010}, along with X-ray light curves from {\it Swift}-XRT and \integral\ (see Sect.~\ref{par:x_rays}).

The historical light curve shows that the gamma-ray activity has increased during the years from the EGRET to \fermi\ era up to factor $\sim$8, thus making \pks\ one of the brightest \fermi\ LAT monitored blazar.

\subsection{X-rays}\label{par:x_rays}            

The \integral\ satellite \citep{Winkler2003} observed  \pks\  in a non-continuous way  starting on 2009 Jan 11, 15:14:28 UT and ending on 2009 Jan 24, 15:02:44  UT (revolutions 763 to 767),  and again between 2010 Jan 17, 14:15:10 UT and 2010 Jan 19, 2010, 04:16:49 UT  (revolution 887).
A 5x5 dither pattern was adopted, so that the total on-source exposure time of the IBIS/ISGRI instrument \citep{Ubertini2003,Lebrun2003} was  405 ks in 2009 and 95 ks in 2010.

Screening, reduction, and analysis of the \integral\ data were performed using the \integral\ Offline Scientific Analysis (OSA) V.~10.1 and IC V. 7.0.2 for calibration, both publicly available through the \integral\ Science Data Center\footnote{http://isdc.unige.ch/index.cgi?Soft+download} 
\citep[ISDC,][]{Courvoisier2003}.  The algorithms implemented in the software are described in \citet{Goldwurm2003} for IBIS, 
\citet{Westergaard2003} for JEM-X, and \citet{Diehl2003} for SPI. The IBIS/ISGRI, SPI, and JEM-X data were accumulated into final images.  For the spectral analysis we used the matrices available in OSA (V. 10.1).

In 2009 IBIS/ISGRI detected \pks\ up to $E = 100$~keV with an average count rate in the coadded image of   $0.39\pm0.05$~counts~s$^{-1}$ in the energy range 20-100 keV. 
The spectrum in this energy range is well fitted by a single power-law $F_E \propto E^{-\Gamma}$  with photon index 
$\Gamma =  1.7\pm0.6$ and the flux  is $(3.4\pm0.9) \times 10^{-11}$  erg~cm$^{-2}$~s$^{-1}$. { The relatively low Galactic absorption, $N_H = 7 \times 10^{20}$~cm$^{-2}$ \citep{Kalberla2005}, is not influential at these energies.} 
The 2010 IBIS/ISGRI average count rate in the coadded image is 
$0.83\pm0.08$~counts~s$^{-1}$ (20-100 keV). The spectrum  has a photon index $\Gamma =  1.4_{-0.5}^{+0.6}$ and a flux of $(6.2\pm1.4) \times 10^{-11}$  erg~cm$^{-2}$~s$^{-1}$.

The IBIS/PICsIT, SPI and JEM-X instruments  on-board \integral\  have not detected the source at either epoch.


%


\pks\  was observed  with  the {\it Swift}  XRT \citep{Burrows2005} in Jan 2009 \citep{Abdo2010c,Dammando2011} and  four times in Jan 2010 (see Tab.~\ref{table:XRTobs}).    The data reduction and analysis of the 2009 data was presented in \citet{Abdo2010c}. The 2010 data were processed following usual procedures as detailed, e.g., in \citet{Dammando2009,Dammando2011}, i.e., the source events were extracted in circular regions centered on the source with radii depending on the source intensity (10--20 pixels, 1 pixel $\sim2.37$~arcsec), while background events were extracted in source-free annular or circular regions. 
Given the low count rate  ($<0.5$ counts~s$^{-1}$),  
we only considered photon counting (PC) data and further selected XRT grades 0--12.  No pile-up correction was required. 
The  data were deabsorbed assuming Galactic HI column density $N_H=7\times10^{20}$ cm$^{-2}$ \citep{Kalberla2005}.  
XRT spectra were extracted for each  observation. Ancillary response files, accounting for different extraction regions, vignetting, and Point-Spread Function (PSF) corrections, were generated with \textsc{xrtmkarf}. We used the spectral redistribution matrices  
in the Calibration Database maintained by HEASARC. All spectra were rebinned with a minimum of 20 counts per energy bin to allow $\chi^2$ fitting within \textsc{XSPEC} (v11.3.2) and
fitted to single power-laws whose indices are reported in Tab.~\ref{table:XRTobs}. 

The RXTE satellite observed \pks\ from 1996 to 2012 with the PCA instrument. 
Count rates in the 2-15~keV band were retrieved from the archive and corrected for the background.  Light curves were produced in two bands (A and B, covering approximately the 2-7.5 keV and 7.5-15 keV, respectively, with some dependence of these bands on mission lifetime).

In panel~(b) of Fig.~\ref{fig:hist.lcurve} we report the sum of the rates in  A and B bands (so called C band, 2-15 keV) and 3-$\sigma$ upper limits, with no correction applied for the negligible neutral hydrogen column density.
In panel~(c) we report the hardness ratio { B/A}.
To improve the statistics we limit our long-term variability analysis to the C-channel only (see Sect.~\ref{sec:multilambda_variability}).


\subsection{Near-infrared-Optical-Ultraviolet}\label{par:optical}            

{\it Swift} UVOT  \citep{Roming2005} data acquired in Jan 2010 were reduced following procedures adopted in  previous  observations  \citep{Dammando2009,Dammando2011}.
Counts-to-magnitudes conversion factors from \citet{Poole2008} and \citet{Breeveld2011} were assumed\footnote{http://swift.gsfc.nasa.gov/analysis/uvot\_digest/zeropts.html}.  The resulting UVOT magnitudes are reported in Tab.~\ref{table:UVOTobs}.
 
We have observed the target in Jan 2009 and Jan 2010 with the StanCam, NOTCam and ALFOSC cameras and optical/near infrared (NIR) filters at the 2.5m NOT located on La Palma.  The data were reduced following standard procedures within the IRAF package \citep{Tody1986,Tody1993}.
A bias subtraction, pixel-to-pixel flat fielding using the normalized internal Halogen
data, and wavelength calibration using internal HeNe-lamp were applied to the raw images.
Aperture photometry was performed with the IRAF\footnote{IRAF is distributed by the National Optical Astronomy Observatory, which is operated by the Association of Universities for Research in Astronomy (AURA) under a cooperative agreement with the National Science Foundation.}   routine APPHOT and calibrated using a standard photometric sequence\footnote{http://www.lsw.uni-heidelberg.de/projects/extragalactic/charts/1510-089.html}.  
Various comparison stars were used depending on filter.   The NOT magnitudes (in Bessel system, except the $i$-band which is in AB system) are reported in Tab.~\ref{table:NOTphotometry} and in Fig.~\ref{fig:hist.lcurve}d.

The 60~cm diameter Rapid Eye Mount (REM) telescope at the ESO site of La Silla observed \pks\ between Apr 2005 and June 2012 in $VRIJHK$ filters. We report in panel~(d) of Fig.~\ref{fig:hist.lcurve} the REM V-band light curve from \citet{Sandrinelli2014a} as well as the V-band archival photometry obtained at the Steward Observatory \citep{Smith2009}.
For both light curves, the relation $V_{Johnson} = V_{AB}+0.044$ is used to convert V-band Johnson-Cousins ($V_{Johnson}$) to AB ($V_{AB}$) magnitudes \citep{Frei1994}.

On 18 Jan 2010  we acquired polarimetry of the source   in several bands with  ALFOSC (exposure time of 100 s).  The  polarized percentages, reported in
Tab.~\ref{table:polarization}, were obtained   using an aperture radius of 2 arcsec for band $B$ and 3~arcsec for bands $z$, $i$, and $V$. 
While the polarization percentage increases with wavelength,  the polarization angle, once 
systematic uncertainties $\sim10\%$ for the latter are taken into account, stays constant within the errors. 

\begin{figure*}[htbp]
\includegraphics[width=0.8\textwidth]{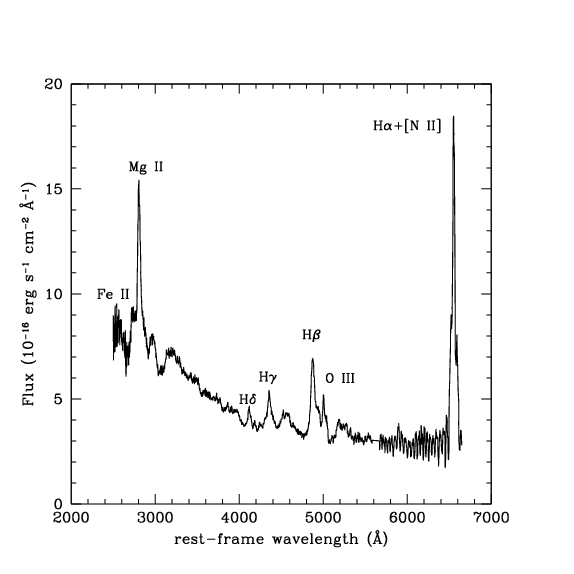}
\caption{Rest frame optical spectrum taken at the NOT with ALFOSC on 22 Feb 2009 and corrected for Galactic extinction ($E_{B-V} = 0.09$).  The most prominent emission lines are labeled.  Some fringing { (removed in the figure)} affects the spectrum long-ward of $\sim$5600 \AA.} 
\label{fig:optical_spectrum_2009}
\end{figure*}

A spectrum of the source was taken on 22  Feb 2009 with ALFOSC  and  reduced following standard procedures within the IRAF package \citep{Tody1986,Tody1993}.  {The data were reduced following standard procedures within the IRAF (overscan subtraction, bias correction, twilight flat fielding). Aperture photometry was done with the IRAF routine APPHOT.}
The spectrum (Fig.~\ref{fig:optical_spectrum_2009}) shows several bright emission lines as well as evidence of a ``small blue bump''  due to the blending of iron lines, often detected in  AGNs in the rest frame range $\sim$2200-4000 \AA\  \citep{Wills1985,Elvis1985}.
The sum of the luminosities of the six most luminous emission lines (H$_\alpha$,  H$_\beta$+OIII, 
H$_\gamma$,  H$_\delta$,  MgII, FeII) yields $L_{lines} = 2.3 \times 10^{44}$~erg~s$^{-1}$. Following \citet{Francis1991},
we correct for the presence of unobserved lines, in particular  the Ly$\alpha$+NV blend, which is expected to contribute significantly to the total line emission, since its observed flux is $0.88 \times10^{-13}$~erg~s$^{-1}$~cm$^{-2}$ \citep{Osmer1994}, i.e., $\sim$50\% brighter than H$_\alpha$ when reddening is taken into account.
We find that the estimated broad line region (BLR) luminosity is $L_{\rm BLR}=5.3 \times 10^{44}$~erg~s$^{-1}$ which is consistent within a factor of $\sim 2$,  and considering its year time scale variability  \citep{Isler2015}, with $L_{\rm BLR}=7.4 \times 10^{44}$~erg~s$^{-1}$ reported by \citet{Celotti1997}.
The estimated $L_{\rm BLR}$ is a factor 7 less than the assumed disk luminosity  (see Sect.~\ref{sec:SEDmodeling}), consistent with a BLR-to-disk covering factor $\sim1/10$ assumed in previous work \citep{Boettcher_Els2016}.

\section{Multiwavelength variability}\label{sec:multilambda_variability}      

\subsection{Variability amplitude}\label{sec:fractional_variability}

The hard X-ray flux (10-50 keV) of  \pks\  varies altogether by nearly a factor of 3 during the period going  from Aug 2006, when {\it Suzaku} observed it to be  $\sim 3.8 \times 10^{-11}$ erg~s$^{-1}$~cm$^{-2}$  \citep{Kataoka2008}, to the first {\it INTEGRAL} IBIS observation  in Jan 2008 \citep[$\sim 1.6 \times 10^{-11}$ erg~s$^{-1}$~cm$^{-2}$, ][]{Barnacka2009},  and our IBIS observations in Jan 2009 ($\sim 2.8 \times 10^{-11}$ erg~s$^{-1}$~cm$^{-2}$) and Jan 2010 ($4.1 \times 10^{-11}$ erg~s$^{-1}$~cm$^{-2}$, see Sect.~\ref{par:x_rays}).  
In Jan 2009 it shows daily variability by about a factor of $\sim$2 during our campaign,  while the soft X-ray (0.3-10 keV) flux varies by at most  30-40\%, and the gamma-ray flux  varies by a factor $\sim$5  (Fig. 1, left).  In Jan 2010, the MeV-GeV flux is  about at the same level as in Jan 2009 and varies with similar amplitude, while the soft X-ray flux varies by a factor of $\sim$2, although on average it is at the same level as in Jan 2009  (Fig. 1, right).    These results indicate a complex behavior in the  X-ray and gamma-ray correlated light curves on time scales from years down to days and
suggest that different components drive the variability at the different  epochs.

We have then considered the multi-wavelength historical light curves (Fig.~\ref{fig:hist.lcurve}) and evaluated the fractional variability at all wavelengths  following \citet{Fossati2000,Vaughan2003}.   In doing so, 
we conservatively discarded the 90\% \fermi\ LAT and 3-$\sigma$ RXTE PCA upper limits.
The results are reported in Tab.~\ref{table:fracrmsvar} and indicate that the variability amplitude   is wavelength-dependent, but not monotonic.   It increases between radio and optical, but at  X-rays is similar to that at radio wavelengths, and it reaches a maximum at gamma-rays.  This is in line with what is observed in the individual epochs \citep{Sandrinelli2014a,Kushwaha2016}, and with the general  behavior of Flat-Spectrum Radio Quasars, where the radio-to-ultraviolet and X-ray-to-gamma-ray parts of the spectrum are dominated by different components, synchrotron emission and inverse Compton scattering, respectively, that are however correlated  \citep{Wehrle1998,Vercellone2011,Pian2011,Tavecchio2013}.  In view of this and of the  lack of a clear pattern in the high energy variability, we have cross-correlated the long-term gamma-ray, X-ray, optical, { and radio} light curves in search of possible delays.

\subsection{Cross-Correlation}\label{sec:cross-correlations} 
\begin{figure*}[hbtc] \centering
\subfloat[]{\includegraphics[width=0.4\textwidth,natwidth=610,natheight=642]{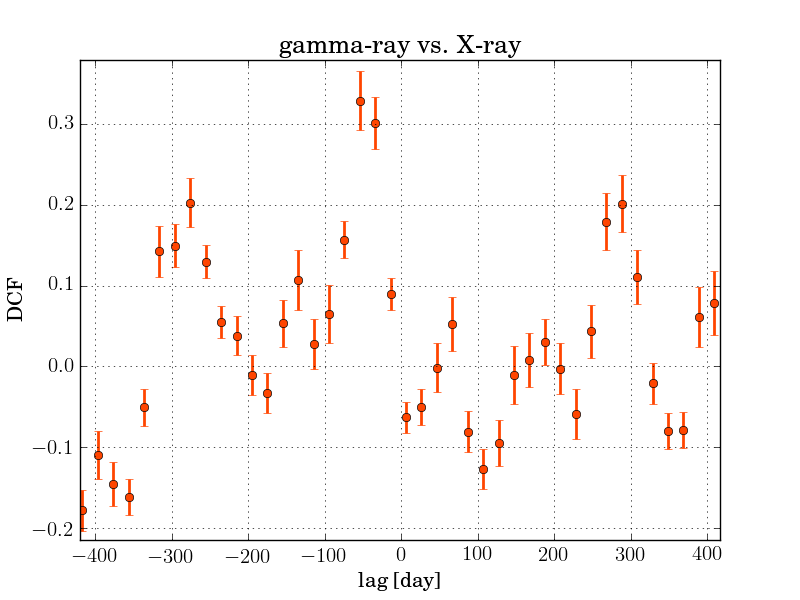}}
\subfloat[]{\includegraphics[width=0.4\textwidth,natwidth=610,natheight=642]{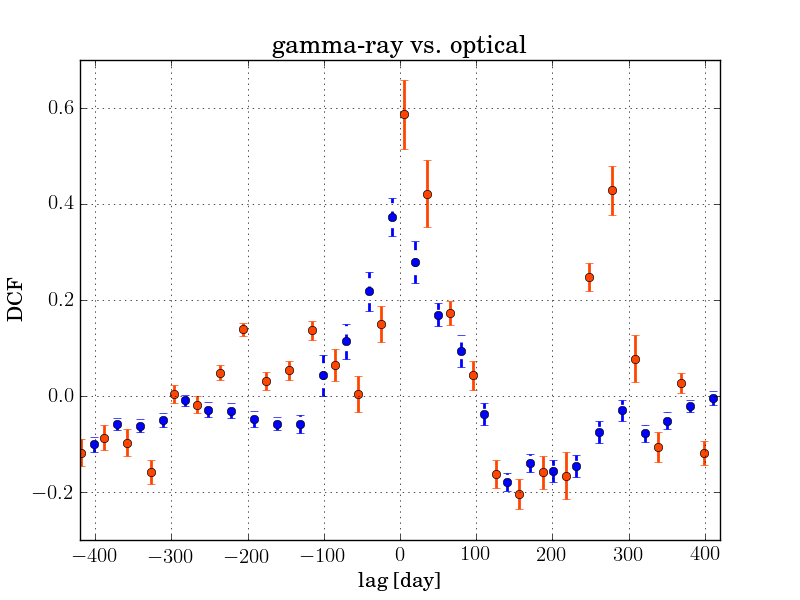}}\\
\subfloat[]{\includegraphics[width=0.4\textwidth,natwidth=610,natheight=642]{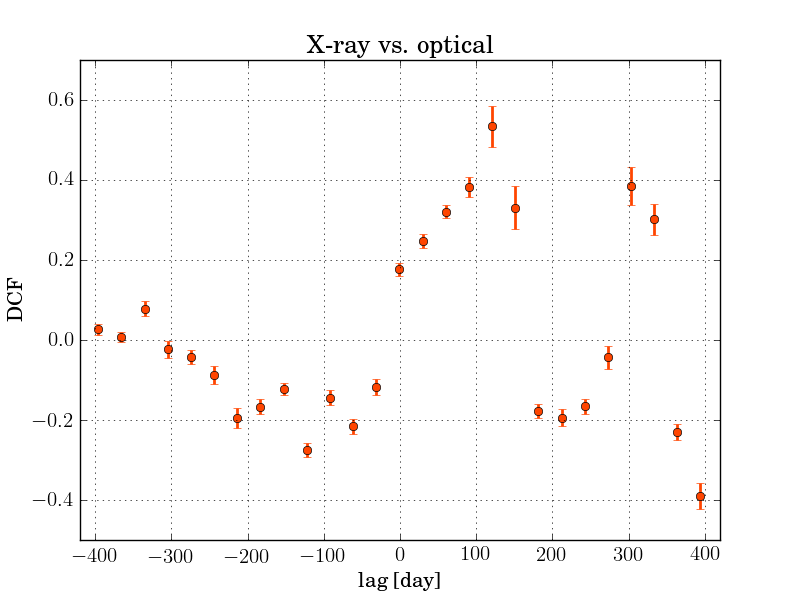}}
\subfloat[]{\includegraphics[width=0.4\textwidth,natwidth=610,natheight=642]{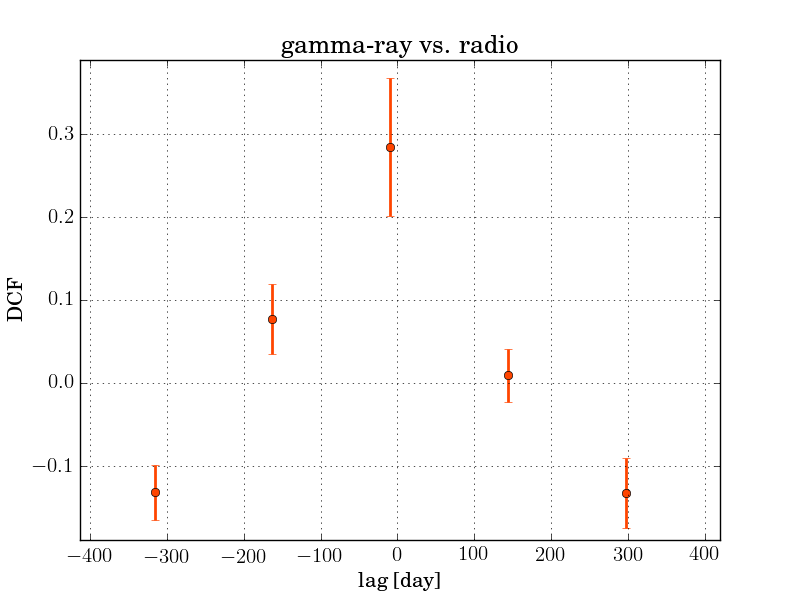}}\\
\subfloat[]{\includegraphics[width=0.4\textwidth,natwidth=610,natheight=642]{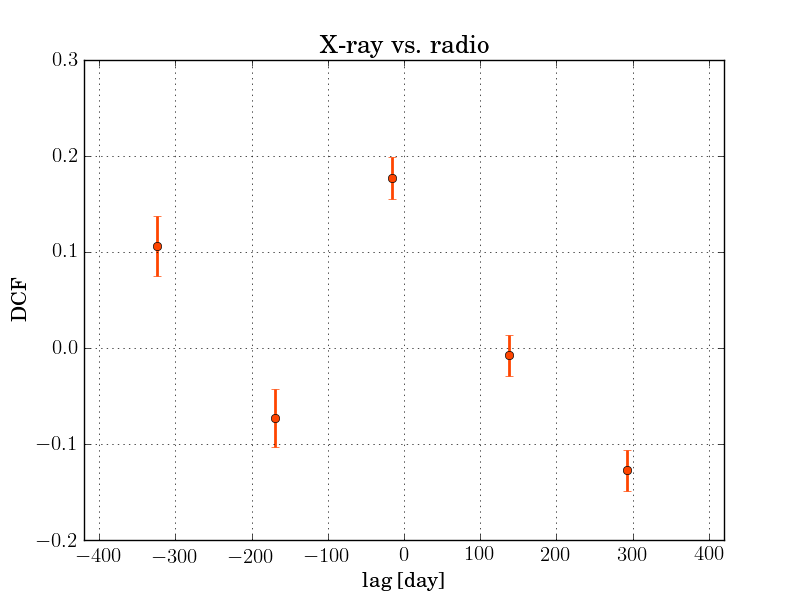}}
\subfloat[]{\includegraphics[width=0.4\textwidth,natwidth=610,natheight=642]{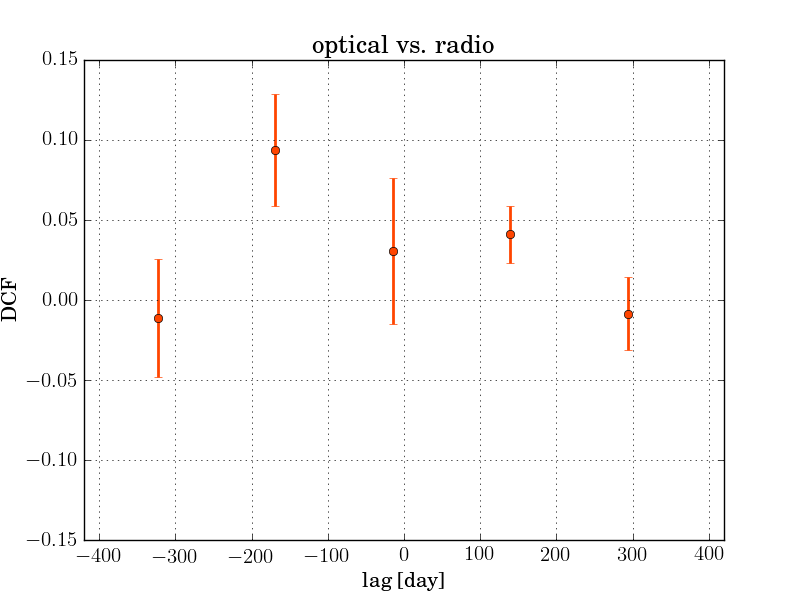}}\\
\caption{ DCF curves between pairs of gamma-ray ({\it Fermi} LAT, 0.1-300~GeV), X-ray
(RXTE PCA, 2-15 keV), optical (Steward Observatory V-band), and radio
(15 GHz) light curves considered in their common time range, i.e.,
between 4411.5 and 5508.1 days from 1 Dec 1996 (orange circles and solid error bars). In
panel (b), we have reported as blue circles and dashed error bars also the DCF obtained by
considering the optical and gamma-ray light curves within their maximum
common time range, i.e., between 4411.5 and 7175.2 days from 1 Dec 1996.
The reported errors are the statistical 1-$\sigma$ uncertainties. The DCF time
bin adopted is equal to 20~days (a), 30~days (b, c), and 150~days (d, e, f) and it is approximately equal to three times the mean time separation between consecutive observations of the light curve with the worse sampling. In each panel positive time lags
correspond to variations of the lower-energy light curve preceding those of the higher-energy light curve.}
\label{fig:DCF}
\end{figure*}

 We have applied the Discrete Correlation Function \citep[DCF,][]{Edelson_Krolik1988}, which is widely used to estimate cross- and auto-correlations of unevenly sampled data of AGNs \citep{Urry1997,Kaspi2000,Raiteri2001,Onken2002,Zhang2002,Abdo2010e,Ackermann2011,Agudo2011}. 
{  We have considered pairwise the gamma-ray ({\it Fermi} LAT, 0.1-300 GeV),  X-ray (RXTE PCA, 2-15 keV), optical (Steward Observatory V-band), and radio
(MOJAVE VLBA, 15 GHz) light curves within their common time range, i.e., between 4411.5 and 5508.1 days from 1 Dec 1996. Within this time range the mean (median) time separations of consecutive observations for the four light curves are $7.6\pm2.5$~days (7.0~days), $2.8\pm5.3$~days (2.0~days), $9.4\pm24.0$~days (1.0~days), and $50.6\pm30.9$~days (54.0~days), respectively. The uncertainties denote the rms dispersion. The DCF time bin was chosen to be $\sim$3 times larger than the average time resolution of the worse sampled light curve and DCF time lags are considered as significant if they are at least 3 times larger than the adopted DCF time bin \citep[see][]{Edelson_Krolik1988,Castignani2014}.
The resulting six DCFs are shown in Fig.~\ref{fig:DCF}. Positive time lags correspond to higher energy photons lagging lower energy photons.

We have also computed the DCF in the maximum time range of the two light curves considered in each panel, but this yields a significantly different result with respect to the above DCFs only for the optical and gamma-ray light curves (Fig.~\ref{fig:DCF}b): the DCF maximum at +300~days time lag disappears. Since the time range considered with the latter case (4411.5 and 7175.2 days from 1 Dec 1996) is larger than in the former case, we deem this result more statistically significant, and therefore we dismiss the 300~days time lag determined in the former case.}

In the gamma-ray vs. optical DCF (Fig.~\ref{fig:DCF}b) marginal correlation at zero time lag is observed.  A  zero time lag between optical and gamma-ray photons { is} consistent with the delays found by  \citet{Abdo2010c} and \citet{Nalewajko2012} over a different time interval (13 days, zero time lag, and 25 days) if one considers that our time-resolution is $\sim$1 week { and $\sim$9~days for the gamma-ray and optical light curves, respectively.}   { Similar time delays consistent with zero between the optical and the gamma-ray photons have been also found other FSRQs such as 3C~454.3 \citep{Bonning2009,Gaur2012,Kushwaha2017}.}

{ The gamma- vs. X-ray DCF (Fig.~\ref{fig:DCF}a) and the X-ray vs optical DCF (Fig.~\ref{fig:DCF}c) suggest that the X-rays lag both the gamma-rays and the optical by an observed time interval between 50 and 300 days. While this may be compatible with the fact that the gamma-rays and optical are produced in the same region by the synchrotron and inverse Compton cooling, respectively, of the same electrons (see also lack of time delay between optical and gamma-rays, Fig.~\ref{fig:DCF}b), the long delay of the X-rays (about 40 to 200 days in rest-frame) with respect to gamma-rays is difficult to reconcile with a physical time scale \citep[see][for discussion of this point]{Hayashida2012}, although it is generally in agreement with the fact that the X-rays may be produced in a more external region with respect to the innermost part of the jet.}

{ The small amplitude of the gamma- vs X-ray DCF makes our results consistent with those of \citet{Abdo2010c}, who did not find any robust evidence of cross-correlation between X- and gamma-rays for PKS~1510-089 at zero time lag during a shorter period (11 months) than examined here. Correlation of X-rays with optical was not reported in previous studies of FSRQs \citep[e.g.,][]{Bonning2009,Gaur2012}. }

{ The poor sampling of the radio light curves prevents us to draw firm conclusions on the corresponding DCFs with emission at shorter wavelengths (Fig.~\ref{fig:DCF}d,e,f). We also checked that our results are independent of the adopted Fermi LAT band and that they do not change significantly if both REM and Steward Observatory V-band light curves are altogether considered.}

{ We also performed the cross-correlation analysis using the $z$-transformed DCF \citep[$z$DCF,][]{Alexander1997,Alexander2013} and found results entirely consistent with those obtained with the DCF.}


\subsection{Search for periodicity}
{ Persistent periodicity in AGNs has never been convincingly detected, the case with the most accurate observational claim being the  optical light curve of the BL Lac object OJ~287 \citep{Sillanpaa1988,Takalo1994,Pihajoki2013,Tang2014a,Tang2014b}. 

Furthermore, if binary systems of SMBHs are present at the center of active galaxies, these may 
produce periodicities, detectable at many wavelengths, on various timescales related to the binary orbital motion.  In view of their role as potential emitters of low frequency gravitational waves, we  are particularly interested in searching for periods in putative SMBHs  that would be typical for binary separations of  $\sim$0.1~pc, i.e., where direct imaging could not distinguish a double source \citep{Komossa2003,Rodriguez2006,Deane2014,Deane2015}, and before the system starts evolving rapidly toward coalescence.   Because of the long time interval covered and the regular sampling, RXTE PCA data are particularly suited for this task.}

Several methods, some of which adopted from areas different from astrophysics 
\citep{Mudelsee2002,Schulz_Mudelsee2002},  have been employed to estimate periodicities of light curves of astrophysical sources and their significance \citep[e.g,][]{vanderKlis1989,Israel_Stella1996,Huang2000a, Huang2000b, Zhou_Sornette2002,Vaughan2005,Vio2010,Kelly2009,Greco2015}  and  AGNs in particular \citep{Gierlinski2008,Lachowicz2009,Mushotzky2011,MaxMoerbeck2014,Hovatta2014,Vanderplas2015,Lu2016,Charisi2016,Connolly2016}.

However, to the best of our knowledge, none of them takes simultaneously into account the following three circumstances:  i) the data are censored {(i.e., upper/lower limits are present)},  ii) they are unevenly sampled, and iii) they are affected by large low-frequency noise.  Furthermore, a model dependent treatment of the noise, gap filling, and/or Monte Carlo simulations are often invoked \citep[see e.g.,][and references therein]{Graham2015a,Greco2015}. The complexity of the high-energy light curves of \pks\ motivated us not to use any of the above  strategies.

{ We apply independent  techniques, namely the DCF and the Lomb-Scargle method to search for possible periodicities and time correlations in both X-ray (2-15~keV) and gamma-ray light curves reported in Fig.~\ref{fig:hist.lcurve}.}
Concerning the gamma-ray emission, the following long-term temporal analyses refer to the 0.1-300~GeV \fermi\ LAT energy range. We checked that our results are independent from the choice of the energy range, i.e., 0.1-0.3 GeV, 0.3-1 GeV, or 1-300~GeV.

\begin{figure*} \centering
\subfloat[]{\includegraphics[width=0.5\textwidth,natwidth=610,natheight=642]{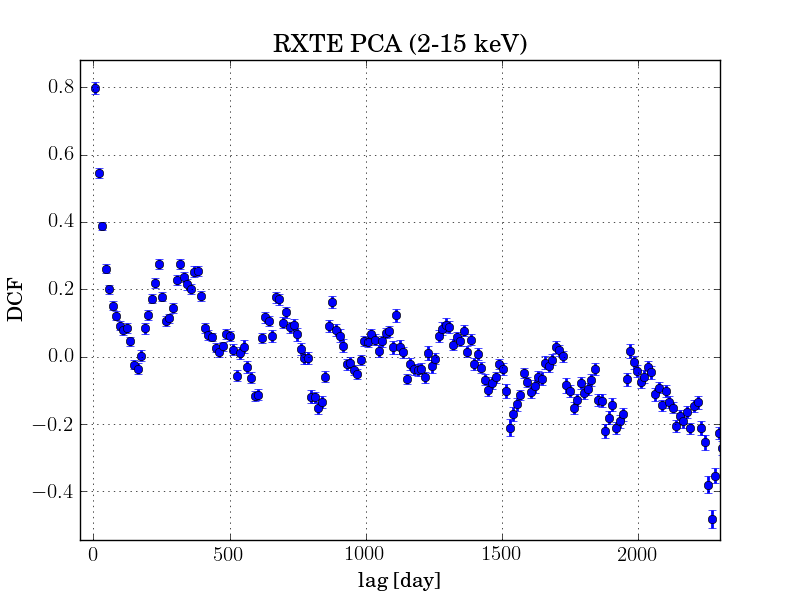}}
\subfloat[]{\includegraphics[width=0.5\textwidth,natwidth=610,natheight=642]{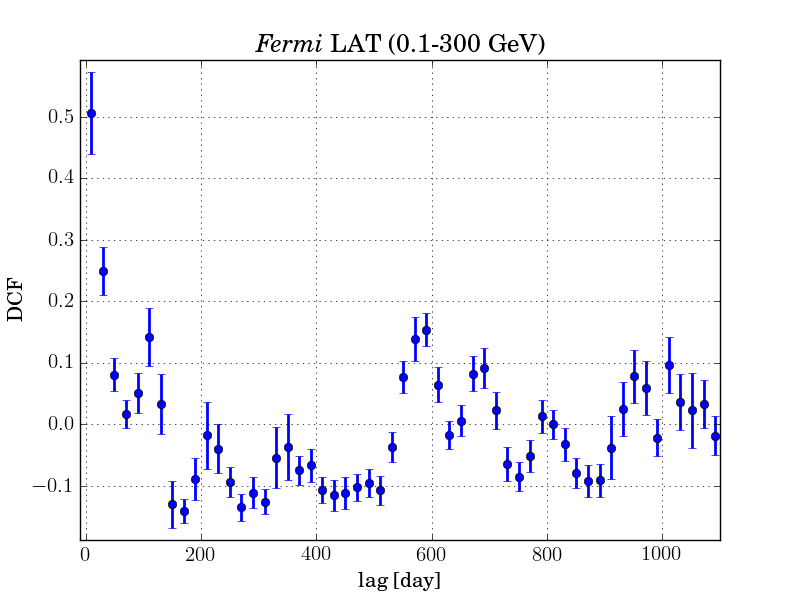}}\\
\caption{Auto-correlation function { (DCF)} of the  X-ray { (RXTE PCA, 2-15 keV, left)} and  gamma-ray { ({\it Fermi} LAT, 0.1-300~GeV, right)} light curves. 
{ DCF time bins of 13 and 20~days are adopted, respectively.} The reported errors are the statistical 1-$\sigma$ uncertainties.
} 
\label{fig:ACF}
\end{figure*}

In Fig.~\ref{fig:ACF} we report the results of  (quasi-)periodicities search in  X- and  gamma-ray light curves of \pks\ using  the DCF analysis.
{ Consistently with Sect.~\ref{sec:cross-correlations} DCF time bins of 13 and 20~days are adopted, respectively.}   
While  no { clear} correlation maxima are seen in X-rays,  two main peaks at time lags of $\sim$100  and  $\sim$600 days  are observed in gamma-rays.   However,  their DCF amplitudes are small and suggest that the significance  of these peaks is very limited.   Owing to the noisy behavior of the DCF curves, which is in turn related to the high red-noise affecting the light curves, we have not attempted to evaluate the significance of the DCF maxima.
\citep[for instance  by means of Monte Carlo simulations, ][]{Litchfield1995,Zhang1999,Castignani2014}.



We  have also  applied to the gamma- and X-ray light  curves the Lomb-Scargle  method \citep{Lomb1976,Scargle1982} that is suited for the studies of time series of astrophysical sources with unevenly spaced data. 
Because of the presence of uneven gaps in the light curve (due to upper limits, that we removed from the analysis), if any quasi-periodicity is present in our light curves, the data likely do not sample all phases equally and the standard Lomb-Scargle method may erroneously estimate the true time period \citep[see e.g., Chapter 10, Sect.~3 of][]{Ivezic2014}. Therefore we have applied the generalized Lomb-Scargle method \citep{Zechmeister_Kurster2009} that addresses these issues. Nevertheless, we checked that our results are substantially unchanged if the standard Lomb-Scargle periodogram is applied instead.

Because of the intrinsic year time scale coverage we are allowed to investigate a frequency domain down to frequencies $\omega=2\pi/T=1.1\times10^{-3}$~day$^{-1}$ and $2.1\times10^{-3}$~day$^{-1}$ for RXTE PCA and {\it Fermi} LAT, respectively. Here $T$ denotes the time window covered by the observations.
We have conservatively chosen a frequency range $2\pi/T\leq \omega \leq  \pi/10\;{\rm day}^{-1}=0.31\;\mathrm{day}^{-1}$. 
The upper bound is well within the pseudo-Nyquist frequency\footnote{If the data were evenly sampled the maximum allowed frequency would be the Nyquist frequency $\omega=\pi/\Delta T$. Since the data are unevenly sampled a pseudo-Nyquist frequency $\omega = \langle\pi/\Delta T\rangle$  can be chosen instead \citep{Debosscher2007}. Nevertheless some studies show that the maximum allowed frequency may be even higher than $\omega = \pi/\Delta T_{min}$ \citep{Eyer_Bartholdi1999}, where $\Delta T_{min}$ is the minimum among the time separations $\Delta T$ \citep[see Chapter~10, Sect.~3 of][for discussion]{Ivezic2014}.} $\omega = \langle\pi/\Delta T\rangle = 0.93$ and 0.45~${\rm day}^{-1}$ for the  RXTE PCA and \fermi\ LAT light curve, respectively. Here $\langle1/\Delta T\rangle$ is the median value of the inverse of the time separation between consecutive observations.

\begin{figure*} \centering
\subfloat{\includegraphics[width=0.5\textwidth,natwidth=610,natheight=642]{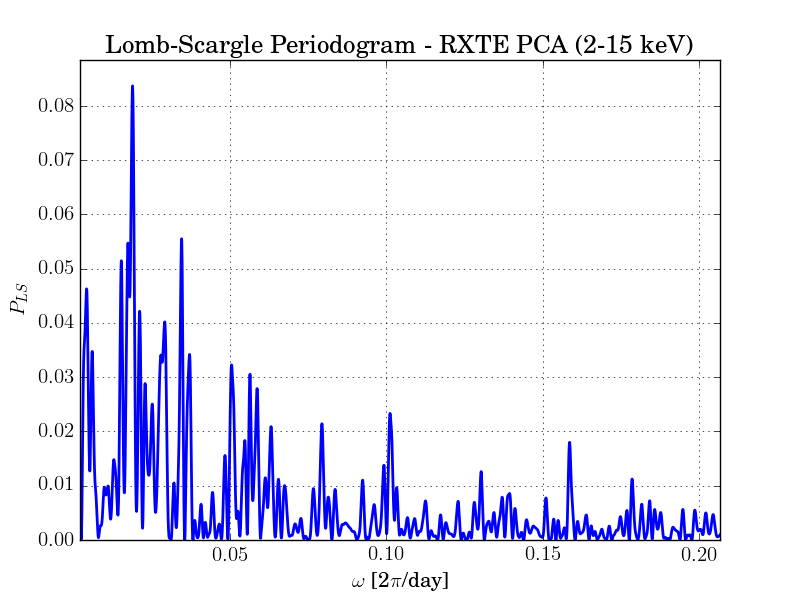}}
\subfloat{\includegraphics[width=0.5\textwidth,natwidth=610,natheight=642]{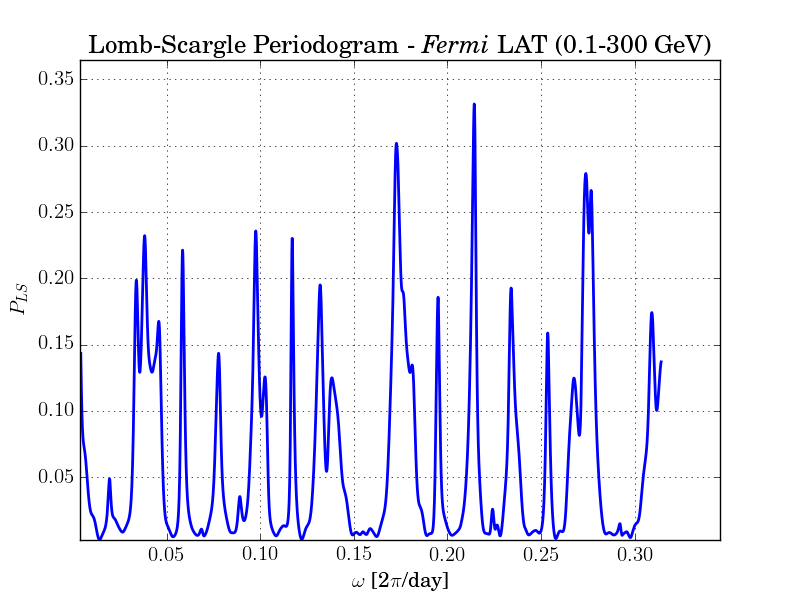}}
\caption{Generalized Lomb-Scargle periodograms \citep{Zechmeister_Kurster2009} for the  X-ray { (RXTE PCA, 2-15 keV, left)} and  gamma-ray { ({\it Fermi} LAT, 0.1-300~GeV, right)} light curves.} 
\label{fig:LS_periodograms}
\end{figure*}

The periodograms for both \fermi\ LAT and RXTE PCA light curves are plotted in Fig.~\ref{fig:LS_periodograms}.
Low-frequency (red) noise is apparent, especially in the RXTE PCA periodogram. 
The periodograms are still very noisy  if a broader frequency range is explored. This is because the low frequencies are affected by red noise and at high frequencies the periodogram flattens to  approximately null values.
Both periodograms present several peaks;  however, 
as for the auto-correlation function (Fig.~\ref{fig:ACF}), the presence of red noise prevents us from estimating their significance using the original prescription reported in \citet[][]{Scargle1982}, which relies on the exclusive presence of white noise in the data.  


Among the peaks present in the X-ray periodogram of Fig.~\ref{fig:LS_periodograms} the { highest 
is the one} associated with a period of $\sim333$~days. 
This period is in agreement with the quasi-periodic optical flux minimum (period of 336$\pm$14~days) and the 336$\pm$15 days radio quasi-periodicity that were found based on data acquired between 1990 and 2005  \citep{Xie2002,Xie2008}.
However, this period is close to the Earth orbital period, it
is not seen in the gamma-ray Lomb-Scargle periodogram, and it is not { significantly } seen with the auto-correlation function (DCF) of either gamma- and X-ray light curves  (Fig.~\ref{fig:ACF}). We conclude that the $\sim$1~yr period is likely spurious.


Our findings are formally consistent with the results of \citet{Sandrinelli2016}, who reported evidence of 115~day quasi-periodicity in 0.1-300~GeV \fermi\ LAT light curve of \pks. Such a value is consistent with the peak we detect at $\omega\simeq0.05$~day$^{-1}$ (Fig. \ref{fig:LS_periodograms}).   However, they did not find the same period in the optical.


\section{Spectral energy distributions}\label{sec:SEDs}            

\subsection{Broad-band spectra construction}  \label{sec:SEDs_panels} 

In Fig.~\ref{fig:SED_multiepoch_1510} we  report the data of our multiwavelength campaigns in Jan 2009 (panel~c) and Jan 2010 (panel~d) along with  archival simultaneous or quasi-simultaneous data corresponding to 2007-2008 (panel~b) and to the epoch of { 2006 (panel~a)}.
{ In panel~(a) we report also archival ROSAT data  \citep{Siebert1996}, that are consistent with the X-ray data of Aug 2006, and CGRO-EGRET data  \citep[$>$100~MeV, Apr 1991 - Nov 1992,][]{Hartman1999}, that  correspond to the average state of the source and are included to provide a hint of the gamma-ray flux, in absence of simultaneous gamma-ray coverage from any satellite in 2006.}
The \fermi\ LAT  data of 2008 and 2009 are from \citet{Abdo2010c}.   We used a photon index $\Gamma=2.3$, from \citet{Ackermann2015a}, to convert LAT count rates to spectral flux densities in Jan 2010.   
In panel~(c) we report also very high energy (VHE) data at few hundred GeV of March 2009, quasi-simultaneous with those at lower frequencies. 



The X-ray and optical data are all corrected for Galactic absorption.
For the conversion of optical  magnitudes to fluxes, we used the photometric zero-points of \citet{Fukugita1995} and \citet{Megessier1995}.   
Before doing so, we  transformed our NOT $K_s$-band magnitudes to $K$-band adopting the relation $K=K'-0.2$~mag \citep{Klose2000,Wainscoat1992} that was inferred from afterglow emission of several gamma-ray bursts (GRBs) assuming a single power law energy distribution for the optical-NIR spectrum, a relativistic synchrotron emission, and the fact that the radiating electrons are in fast-cooling regime, i.e., the characteristic cooling time is shorter than the shock propagation time. Since the $K-$, $K'-$, and $K_s-$band wavelengths satisfy the relations $\lambda_{K}>\lambda_{K_s}\gtrsim\lambda_{K'}$ {we have applied the relation above replacing $K'$ with $K_s$.}

In Fig.~\ref{fig:SED_multiepoch_1510}d we report the polarized NOT  spectrum, where NOT fluxes have been multiplied by the polarization fractions reported in Tab.~\ref{table:polarization} and then by a constant factor of 50. The polarized spectrum, which is  bluer than the unpolarized broad-band optical spectrum, presumably traces only
the synchrotron component (the thermal component being usually unpolarized) and guides its identification for the modeling (see Sect.~\ref{sec:SEDmodeling}).

Because of the seemingly "incoherent" character of multi-wavelength variability (i.e., no clear correlation among the multi-wavelength light curves, see Sect.~\ref{sec:multilambda_variability}), and owing to the lack of strict simultaneity at the first and second epochs under study (panels a and b),  it is not straightforward to define a quiescent vs flaring state, unlike in \citet{Abdo2010c}, where the regular gamma-ray monitoring allows a neat definition of 4 flaring episodes.  We consider here  as flaring states those where the gamma-ray activity was most dramatic and gamma-ray flux  detected at its highest, i.e., Jan 2009 and Jan 2010.    During these 2 states, the hard X-ray  flux was also relatively high, i.e., about at the same level as detected by {\it Suzaku} in 2006 (panel a), while the optical fluxes differ by a factor of $\sim$2.
The 2007-2008 epoch (panel b) appears to represent the lowest multi-wavelength state, although we caution that the hard X-rays are not simultaneous (by 2 months or more) with optical and gamma-ray data. 

Previous modeling of the \pks\ multi-wavelength energy distribution has focused on single epochs and aimed at defining the interplay of different emission components during  the evolution of a flare.    In the next section we compare our four SEDs with a blazar model in order to characterize the long-term multi-wavelength variability and to identify the main parameters that are responsible for year time scale variations in this blazar.

\subsection{Spectral model}\label{sec:SEDmodeling}       

We have applied  the blazar leptonic model described in \citet{Ghisellini_Tavecchio2009}
to the four SEDs of  Fig.~\ref{fig:SED_multiepoch_1510}.  

The model envisages thermal components from the accretion disc, a dusty torus, and a corona emitting mainly in the optical-ultraviolet,  far-infrared (FIR), and at X-rays, respectively, as well as synchrotron radiation at radio-to-ultraviolet  frequencies  from a homogeneous region,  peaking in the FIR (dot-dashed curve in all panels of Fig.~\ref{fig:SED_multiepoch_1510}).  However, since most of the radio emission is produced at significantly larger distances from the central nucleus than the higher frequency emission, our model systematically underestimates it 
\citep[this is clearly seen in Fig.~\ref{fig:SED_multiepoch_1510}a,b; see also discussion of this point in ][]{Ghisellini2011}.  Following \citet{Kataoka2008} we have associated the optical-ultraviolet  spectrum with a standard   \citep[i.e., optically thin and geometrically thick, ][]{Shakura_Sunyaev1973} accretion disk
of constant  luminosity $L_d  = 5.9\times10^{45}$~erg~s$^{-1}$.   This is also responsible for powering the luminous optical emission lines (Fig.~\ref{fig:optical_spectrum_2009}).  
Its corona emits at X-rays with an assumed  
luminosity $\sim$ 30\%\:$L_{d}$ and
spectrum $\sim \nu^{-1}exp(-h\nu/150 keV)$.    
The assumed  BLR radius  is $R_{BLR}$ = 2.4 $\times$ 10$^{17}$~cm, on the basis of relations based on reverberation mapping \citep[e.g.,][]{Kaspi2005}.
Inverse Compton scattering occurs off both synchrotron photons (synchrotron self-Compton, SSC) and photons external to the jet (EC), primarily associated with the disk and BLR, and produces the  luminous  X- and gamma-ray component peaking in the MeV-GeV range.

{ We modeled Mar 2009  data from HESS at a few hundred GeV with an additional component with respect to the external Compton scattering used to reproduce the Fermi-LAT data (Fig.~\ref{fig:SED_multiepoch_1510}c). This is because Inverse Compton scattering off BLR photons (with comoving energy density 6.6$\times10^{-5}$~erg~cm$^{-3}$) does not reproduce the VHE emission (the pair-production opacity is dramatic). Advocating FIR photons (with comoving energy density 4.3$\times10^{-2}$~erg~cm$^{-3}$) from a torus located at $\sim0.2$~pc from the nucleus as  a source for external Compton  scattering  we obtain a satisfactory fit of the VHE part of the spectrum.}



The SED model parameters  are reported in Tab.~\ref{table:modelparams} and the corresponding curves are shown in Fig.~\ref{fig:SED_multiepoch_1510}.  In panel (d) are plotted the model curves for all states.
The jet viewing angle $\theta_v$ is 3$^\circ$ for all states, so that
the Doppler factor $\delta$ varies from $\sim$18 (when $\Gamma=13$) to $\sim$19 ($\Gamma=16$).
These parameters are in general good agreement with the modeling results obtained for Mar 2008 and Jan 2009 by \citet{Dammando2009,Abdo2010b,Abdo2010c,Boettcher2013}.  {We note that our best-fit value for the electron distribution cooling break, $\gamma_b$, in Jan 2009 is a factor $\sim$7 higher than found in \citet{Ghisellini2010} for the period June - Aug 2008.}

Among the model parameters  (Tab.~\ref{table:modelparams}),  the total power injected into the jet, $P'_i$,  and  the random Lorentz factor, $\gamma_b$, corresponding to the electron distribution cooling break  stand out  as the most clearly variable and are thus likely responsible for most of the variability at the four epochs under exam.  This
suggests a correlation of the injected power with enhanced activity of the acceleration mechanism
\citep[see also ][]{Nalewajko2012}.    In particular, 
$P'_i$ is highest at the latest epoch (Jan 2010),  when also the power distributed in the 4 channels of energy output (radiation, Poynting flux, electron and proton power) is highest.

{We stress that, although some degeneracy among { the} best fit parameters exists, data effectively constrain several physical quantities.
{ For example, the size of the emitting region is estimated from the shortest} typical variability time scale; the ratio of the comoving energy density associated with the magnetic field in the jet to that associated with the radiation field is constrained by the relative height of the two main peaks in the SED; the random Lorentz factor, $\gamma_b$, by the distance of the two peaks; the maximum random Lorentz factor, $\gamma_{\rm max}$, by the maximum energy in gamma-rays; the incidence of the thermal component associated with the accretion on to the AGN by the optical-UV bump in the SED. Therefore, although affected by some uncertainty, the variations in the parameters among different states of the source are significant.}

We used the disk rest-frame peak frequency $\nu_{\rm peak}\simeq3.6\times10^{15}$~Hz and the disk luminosity adopted in the model, as well as the corresponding accretion rate $\dot{M}\simeq1~M_\sun$~yr$^{-1}$ to estimate a black hole mass $M_{\rm BH}=2.4\times10^{8}$~M$_\odot$ \citep[Eq.~5 of][]{Castignani2013}, where {an accretion efficiency $\eta=\frac{L_d}{\dot{M}c^2}=0.1$ is assumed and $c$ is the speed of light.} The mass estimate is compatible  with previous estimates, $(2.0-9.1)\times10^{8}~M_\odot$ \citep[][]{Xie2005,Abdo2010c,Liu_Bai2015},  within typical $\sim0.4-0.5$~dex statistical uncertainties \citep{VestergaardPeterson2006,Park2012,Castignani2013} associated with black hole mass estimates.

\section{Conclusions}\label{sec:conclusions}     

\pks,  one of the most powerful gamma-ray blazars continuously monitored by \fermi, was observed by
\integral\  in Jan 2009 and Jan 2010 in outburst.   
The \integral\ IBIS data and simultaneous  {\it Swift} XRT, UVOT, and optical NOT data are presented.  We have studied the multi-wavelength variability of the source based on these data, complemented by archival  CGRO-EGRET,  {\it Fermi} LAT, and AGILE GRID observations in gamma-rays and  published X-ray, optical, and radio data at prior epochs.  The historical variability of this source reflects the  origin  of the broad-band spectrum,  wherein  a population of highly relativistic particles first radiate via synchrotron process at
radio-to-ultraviolet wavelengths, so that the variability amplitude increases monotonically over this spectral band, and then at X- and gamma-rays via inverse Compton scattering off the synchrotron  photons and external photon fields, which causes the variability amplitude to increase with energy in a similar way as the synchrotron.   However, the variations in different bands  do not correlate in a simple way and the time delays that emerge from a quantitative cross-correlation analysis  of the light curves with the DCF method  { (up to about $\sim$200 days in rest frame)} have no straightforward physical meaning.   This typically unpredictable character of blazar  variability occasionally has "orphan flares" as a consequence, i.e., high energy (gamma-rays) outbursts with no simultaneous or quasi-simultaneous (i.e., within a few weeks) counterpart at lower frequencies \citep{Krawczynski2004,Nalewajko2012,Wehrle2012,MacDonald2015}. 

A search  for periodicity in the archival RXTE PCA  and \fermi\ LAT light curves
yields similarly inconclusive  results: the auto-correlation function (DCF) and the Lomb-Scargle  periodogram return the suggestion of  a number of { recurrent} time scales and quasi-periodicities at X- and gamma-rays, whose significance is however difficult to assess, because of red noise affecting the light curves, which  makes probability evaluation impractical.   It must be noted that the appearance of these period candidates only in individual, and not all, bands undermines their authenticity.  Moreover, the results of various methods are not unanimous.
We conclude that there is no robust evidence for (quasi-)periodicities in the high-energy light curves of \pks, at least up to the few-year time scales that we can probe with our light curves.

\begin{figure*}[htbc]  
\begin{center}
\includegraphics[width=0.9\textwidth]{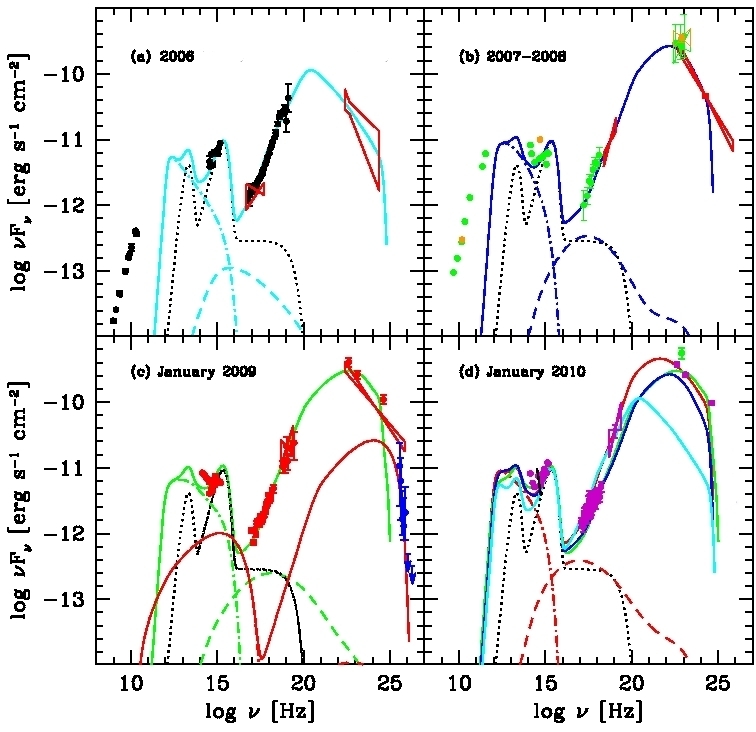}
\caption{\small Simultaneous or quasi-simultaneous multiwavelength spectra of \pks\ at different epochs in rest-frame. Soft X-ray data and ultraviolet-to-NIR data were corrected for neutral hydrogen absorption ($N_H = 7 \times 10^{20}$~cm$^{-2}$) and dust extinction ($E_{B-V} = 0.09$) in our Galaxy, respectively. 
(a) The filled black circles are from \citet{Kataoka2008} and refer to August 2006.  Red bow-ties are archival data from ROSAT \citep{Siebert1996}, and CGRO-EGRET  \citep[$>$100~MeV, Apr 1991 - Nov 1992,][]{Hartman1999}. 
(b) Orange symbols are AGILE-GRID and GASP data obtained during 27 Aug - 1  Sep 2007  \citep{Pucella2008}. Light green symbols are AGILE-GRID, GASP, {\it Swift}/UVOT, and {\it Swift}/XRT data of Mar 2008  \citep{Dammando2009}.  The red bow-tie at hard X-ray frequencies corresponds to the \integral\  spectrum  in Jan 2008 \citep{Barnacka2009}. The red bow-tie at gamma-ray frequencies corresponds to \fermi\ LAT observations during Aug - Oct 2008 \citep{Abdo2009}.
(c) Red symbols refer to optical data from NOT
(13 and 18 Jan 2009);  {\it Swift} XRT and UVOT    \citep[10-25 Jan 2009,][]{Abdo2010c};   \integral\ IBIS-ISGRI (13-24 Jan  2009);  \fermi\ LAT data (10 Jan -- 1 Feb 2009).
Blue symbols are HESS  observations in  Mar 2009 \citep[][]{Abramowski2013};   upper limits at 3-$\sigma$ confidence for the two highest-energy bands are reported.
(d) Magenta symbols refer to optical data from NOT (23 Jan 2010) and {\it Swift}/UVOT (15, 17, and 19 Jan 2010); {\it Swift} XRT  (15-19 Jan 2010),  \integral\  IBIS-ISGRI  (17-19 Jan 2010); and \fermi\ LAT (26 Dec 2009 - 23 Jan 2010).
The green symbol is the $E>100$~MeV gamma-ray observation from AGILE \citep[11-13 Jan 2010, ][]{Striani2010}. 
The black solid segment shows the polarized NOT spectrum of 18 Jan 2010. Fluxes have been multiplied by a constant factor of 50.  
In each panel, we report the curves obtained from a fit of the SEDs with the blazar model \citep{Ghisellini_Tavecchio2009}.  The black dotted curve represents a thermal component that is assumed to be constant in time and consisting of radiation from an accretion disk (optical-ultraviolet), a torus (FIR)  and a corona (X-rays, with an exponential cutoff at $\sim$100 keV).
The dot-dashed curve shows the synchrotron emission component.
The dashed curve represents first and second order synchrotron self-Compton process.  Most of the inverse Compton scattering occurs off external BLR photons (this individual component is not shown).  The sum of the thermal  and non-thermal components is shown as a single curve in each panel: cyan in (a), blue in (b), green in (c) and red in (d), where we report also the best-fit curves for the 3 previous epochs.
The quasi-simultaneous VHE emission of Mar~2009 is modeled independently by External Compton of infrared photons of the torus. The sum of this component and the Synchrotron emission is shown in (c) as solid red curve.}
\label{fig:SED_multiepoch_1510}
\end{center}
\end{figure*}

We assembled  spectral energy distributions of \pks\  centered at the epochs  of hard X-ray observations by {\it Suzaku} (Aug 2006) and \integral\ (Jan 2008, Jan 2009, and Jan 2010), using simultaneous radio-to-gamma data  where available or only { partially simultaneous data (2006, Fig.~\ref{fig:SED_multiepoch_1510}a; 2007-2008, Fig.~\ref{fig:SED_multiepoch_1510}b).}   Through a model for blazar multi-wavelength emission { in a homogeneous region} we evaluated the main parameters that are responsible for driving the variability from epoch to epoch over the years, whereas previous work had only focused on single epochs or flaring episodes of this source.
The largest variations are estimated to occur in the total power injected into the jet and the random Lorentz factor of the electron distribution cooling break. This suggests a correlation of the injected power with enhanced activity of the acceleration mechanism.   We find that the dissipation radius is much smaller than the BLR size, suggesting that the flares must occur well within the BLR.
{ However we note that the model cannot account for emission at a few hundred GeV if this occurs in the same region.  It must be produced in an outer region, about 0.2~pc from the AGN, to avoid suppression by the abundant optical-UV BLR photons (via pair-production).}

Modeling  of broad-band blazar SEDs over a range of time scales is crucial to map the global variability into  the behavior of the central engine.  Accurate time analysis of regularly sampled   multiwavelength light curves adds to  the broad-band spectral diagnostics and helps  in the identification of periodicities.  These can be the signatures of supermassive black hole binary systems,  possibly detectable during their final merger phase  by the upcoming gravitational wave space interferometer eLISA.





\begin{table*}
\caption{{\it Swift} XRT  observations in Jan 2010.} 
\label{table:XRTobs}
\centering                       
\begin{tabular}{cccccc}      
\hline\hline              
[1]  &  [2] &  [3]  &  [4]  &  [5] &  [6]  \\
\hline                 
%
%
2010 Jan 15, 04:19 & 2010 Jan 15 11:58 & 3907 & 0.137 &  $8.15\pm0.66$ & $1.57\pm  0.11$  \\   
2010 Jan 17, 07:47 & 2010 Jan 17 13:31 & 3544 & 0.137 &  $8.44\pm0.69$ & $1.58 \pm 0.11$  \\   
2010 Jan 19, 09:26 & 2010 Jan 19 13:51 & 3998 & 0.162 &  $9.86\pm0.69$ & $1.59 \pm 0.11$  \\   
2010 Jan 30, 01:01 & 2010 Jan 30 22:56 & 4125 & 0.219 &  $15.14\pm0.11$ & $1.30 \pm 0.09$ \\   
\hline 

\noalign{\smallskip}
\multicolumn{6}{l}{[1] Start Time (UT);}\\
\multicolumn{6}{l}{[2] End Time (UT);}\\
\multicolumn{6}{l}{[3] Exposure Time (s);} \\
\multicolumn{6}{l}{[4] Count rate (c/s);} \\
\multicolumn{6}{l}{[5] Flux (0.3-10 keV) in units of $10^{-12}$ erg~s$^{-1}$~cm$^{-2}$, corrected for Galactic absorption $N_H=7\times10^{20}$ cm$^{-2}$.} \\
\multicolumn{6}{l}{\hspace{0.4cm} Uncertainties are at 90\% confidence.} \\
\multicolumn{6}{l}{[6] Photon index ($F_E \propto E^{-\Gamma}$).} \\

\end{tabular}                               
\end{table*}


\begin{table*}
\caption{{\it Swift} UVOT photometry in  Jan 2010.} 
\label{table:UVOTobs}
\centering                       
{\small
\begin{tabular}{ccccccc}      
\hline\hline              
[1]   &     [2]  &  [3] &  [4] & [5] & [6] & [7]\\
Time  &      v   &  b   &  u   &  uvw1 & uvm2 & uvw2\\
\hline                 
2010 Jan 15, 04:19-11:58 &    $16.51 \pm 0.06$ & $16.69 \pm 0.04$ & $16.87 \pm 0.04$ & $17.55 \pm 0.04$ & $17.66 \pm 0.05$ & $17.72 \pm 0.04$ \\ 
2010 Jan 17, 07:47-13:31 &    $16.64 \pm 0.06$ & $16.80 \pm 0.04$ & $16.98 \pm 0.04$ & $17.67 \pm 0.04$ & $17.78 \pm 0.05$ & $17.78 \pm 0.04$ \\ 
2010 Jan 19, 09:26-13:51 &    $16.72 \pm 0.06$ & $16.80 \pm 0.04$ & $16.99 \pm 0.04$ & $17.65 \pm 0.04$ & $17.74 \pm 0.05$ & $17.82 \pm 0.04$ \\ 
2010 Jan 28, 23:31-23:38 &     $16.98 \pm 0.26$ & $16.65 \pm 0.10$ & $16.96 \pm 0.08$ & $17.55 \pm 0.08$ & $17.57 \pm 0.08$ & $17.78 \pm 0.06$ \\ 
2010 Jan 30, 01:01-22:56 &    $16.73 \pm 0.06$ & $16.81 \pm 0.04$ & $17.02 \pm 0.04$ & $17.79 \pm 0.04$ & $17.84 \pm 0.05$ & $18.01 \pm 0.04$ \\ 
\hline 
\noalign{\smallskip}
\multicolumn{7}{l}{\normalsize [1] Observation period;}\\
\multicolumn{7}{l}{\normalsize [2-7] Apparent { AB} magnitudes { (not corrected for Galactic absorption)} and  uncertainties.}\\
\end{tabular}}                               
\end{table*}

\begin{table}
\centering                       
\caption{NOT photometry.}          
\label{table:NOTphotometry}      
\begin{tabular}{ccc}      
\hline\hline              
UT & Filter & Magnitude$^a$  \\    
\hline                 
2009 Jan 13, 06:03:23  &  $K_s$ & $12.61 \pm  0.04$  \\
2009 Jan 13, 06:10:34  &  $H$   & $13.58 \pm  0.04$  \\
2009 Jan 13, 06:19:16  &  $J$   &  $14.43 \pm  0.04$  \\
2009 Jan 13, 06:31:18  &  $U$     & $16.25  \pm 0.06$ \\
2009 Jan 13, 06:38:15  &  $B$     & $16.91  \pm  0.06$ \\
2009 Jan 13, 06:39:46  &  $V$    & $16.56  \pm  0.06$ \\ 
2009 Jan 13, 06:41:17  &  $R$     & $16.10  \pm  0.06$ \\
2009 Jan 13, 06:44:12  &  $I$     &  $15.60  \pm 0.06$  \\
2009 Jan 18, 06:18:05  &  $I$       & $16.02 \pm  0.05$ \\
2009 Jan 18, 06:23:23  &  $V$   &  $16.91 \pm 0.05$  \\
2009 Jan 18, 06:26:19  &  $R$   &  $16.42 \pm 0.05$  \\
2009 Jan 18, 06:29:22  &  $B$     &  $17.14 \pm 0.05$  \\
2009 Jan 18, 06:34:06  &   $U$   &  $16.48 \pm 0.05$   \\
\hline    
\hline                 
2010 Jan 15, 06:54:00   &  $i$    & $15.72\pm0.05$ \\
2010 Jan 15, 06:57:46   &  $R$   & $16.16\pm0.05$ \\
2010 Jan 15, 06:57:46   &  $B$    & $16.80\pm0.06$  \\
2010 Jan 15, 07:01:11   &  $V$   &  $16.57\pm0.05$  \\
2010 Jan 15, 07:07:19   &  $U$    & $16.23\pm0.07$  \\
2010 Jan 18, 07:09:14   &  $K_s$  &  $12.90\pm0.04$  \\
2010 Jan 18, 07:15:01   &  $H$    & $13.89\pm0.04$ \\
2010 Jan 18, 07:22:01   &  $J$     & $14.79\pm0.04$ \\ 
2010 Jan 22, 07:04:03   &  $U$  &  $16.34\pm0.07$  \\
2010 Jan 22, 07:08:50   &  $B$ &  $16.87\pm0.05$  \\
2010 Jan 22, 07:12:17   &  $V$ &  $16.71\pm0.05$  \\
2010 Jan 22, 07:14:54   &  $R$  &  $16.32\pm0.05$  \\
2010 Jan 22, 07:17:20   &  $i$   &  $15.94\pm0.05$  \\
2010 Jan 23, 05:54:11    &  $J$   &  $14.77\pm0.05$  \\
2010 Jan 23, 06:01:20    &  $H$   &  $13.89\pm0.05$  \\
2010 Jan 23, 06:08:45     &  $K_s$  &  $12.84\pm0.05$  \\
2010 Jan 23, 06:17:19    &  $i$  &  $15.88\pm0.05$  \\
2010 Jan 23, 06:19:31    &  $R$  &  $16.32\pm0.05$  \\
2010 Jan 23, 06:21:39    &  $V$  &  $16.67\pm0.06$  \\
2010 Jan 23, 06:24:39    &  $B$  &  $16.91\pm0.06$  \\
2010 Jan 23, 07:07:38    &  $U$  &  $16.27\pm0.06$  \\
\hline
\noalign{\smallskip}
\multicolumn{3}{l}{$^a$ { NOT magnitudes in Bessel system, except the}}\\ 
\multicolumn{3}{l}{{  $i$-band which is in AB system.} Magnitudes are}\\
\multicolumn{3}{l}{not corrected for Galactic extinction.}\\
\end{tabular}                               
\end{table}


\begin{table*}
\caption{NOT polarization measurements of 18 Jan 2010.} 
\label{table:polarization}
\centering    
\begin{tabular}{ccccc}      
\hline\hline              
[1]  &  [2] &  [3]  &  [4]  &  [5] \\
\hline                 
z &  $2.8\pm0.4$ &   $137\pm4$ &  $+0.3\pm0.4$  & $-2.8\pm0.4$ \\
i &  $2.4\pm0.3$ &   $155\pm3$ &  $+1.5\pm0.3$  & $-1.8\pm0.3$ \\  
V &  $1.4\pm0.3$ &   $144\pm6$ &  $+0.4\pm0.4$  & $-1.3\pm0.4$ \\
B &  $1.2\pm0.3$  &  $140\pm7$ &  $+0.3\pm0.3$  & $-1.1\pm0.3$ \\     
\hline 
\noalign{\smallskip}
\multicolumn{5}{l}{[1] Band.}\\
\multicolumn{5}{l}{[2] Polarization percentage.}\\
\multicolumn{5}{l}{[3] Polarization angle (degree). Errors refer to statistical uncertainties.} \\
\multicolumn{5}{l}{\hspace{0.4cm} Systematic uncertainties are $\sim10\%$.}\\
\multicolumn{5}{l}{[4] Stokes parameter {\it Q}  ($\%$).} \\
\multicolumn{5}{l}{[5] Stokes parameter {\it U}  ($\%$).} \\                                       
\end{tabular}                               
\end{table*}

\begin{table*}
\caption{Fractional rms variability amplitude for the historical light curves.} 
\label{table:fracrmsvar}
\centering                       
\begin{tabular}{ccc}      
\hline\hline              
Instrument & Band   & $F_{var}^a$ \\
\hline                 
VLBA                 &  15 GHz                    &  $0.482\pm0.001$      \\    
REM                  &  $V$                    &  $0.53\pm0.02$        \\    
Steward Observatory  &  $V$                    &  $0.647  \pm 0.002$   \\    
RXTE PCA                 &  2-15 keV            &  $0.236 \pm 0.003$   \\    
{\it Fermi}-LAT      &  0.1-300 GeV          &  $1.052  \pm 0.005$    \\   
\hline
\noalign{\smallskip}
\multicolumn{3}{l}{$^a$  Defined as in  \citet{Fossati2000,Vaughan2003}. }\\
\end{tabular}                        
\end{table*}

\begin{table*} 
\centering
\begin{tabular}{llllllllllllll}
\hline
\hline
panel  &$R_{\rm diss}$  &$P^\prime_{\rm i}$  &$B$ &$\Gamma$ 
   &$\gamma_{\rm b}$ &$\gamma_{\rm max}$ &$s_1$  &$s_2$  &$\log P_{\rm r}$ &$\log P_{\rm B}$  &$\log P_{\rm e}$
  &$\log P_{\rm p}$  \\ 
~     &[1] &[3] &[4] &[5] &[6] &[7] &[8] &[9] &[10] &[11]  &[12] &[13]   \\  %
\hline   
a  &72 (800)  &8e--3  &3.8  &13  &10  &4e3   &1    &2.7 &44.7 &44.7 &44.7 &47.2 \\ 
b  &86 (950)  &8e--3  &2.9  &16  &300 &4e3   &1    &3.1 &45.3 &44.8 &44.6 &46.8 \\ 
c  &81 (900)  &7e--3  &2.2  &15  &1e3 &4e3   &1.3  &3   &45.2 &44.4 &44.4 &46.7 \\ 
d  &90 (1e3)  &0.014  &2.5  &15  &200 &3e3   &1.3  &2.6 &45.5 &44.7 &44.9 &47.3 \\ 
\hline
c (VHE) &720 (8e3)  &5e--4  &0.25  &13  &4e4 &8e4  &1.4 &2.7  &44.4 &44.3 &43.1 &44.5 \\ 
\hline
\hline 
\end{tabular}
\vskip 0.4 true cm
\caption{List of parameters used to construct the theoretical SED (Col [2] to Col. [9]) and derived 
jet powers (Col. [10] to [13]).
Col. [2]: dissipation radius in units of $10^{15}$ cm and (in parenthesis) in units of Schwarzschild radii;
Col. [3]: power injected in the blob calculated in the comoving frame, in units of $10^{45}$ erg s$^{-1}$; 
Col. [4]: magnetic field in Gauss;
Col. [5]: bulk Lorentz factor at $R_{\rm diss}$;
Col. [6] and [7]: break and maximum random Lorentz factors of the injected electrons;
Col. [8], [9]: slopes of the injected electron distribution [$Q(\gamma)$] below and above $\gamma_{\rm b}$.
Col. [10], [11], [12], [13]:
Logarithm of the jet power in the form of radiation $(P_{\rm r})$, 
Poynting flux $(P_{\rm B}$,
bulk motion of electrons $(P_{\rm e}$
and protons ($P_{\rm p}$, assuming one proton
per emitting electron). Powers are in erg s$^{-1}$.
The black hole mass is assumed to be $M=3\times 10^8 M_\odot$. The disk luminosity is assumed to be
always the same and equal to $L_{\rm d}=5.9\times 10^{45}$ erg s$^{-1}$ (equivalent to 0.13 $L_{\rm Edd}$.
The radius of the broad line region is assumed to be $R_{\rm BLR}=2.4\times 10^{17}$ cm.
The total X--ray corona luminosity is assumed to be 30 per cent of $L_{\rm d}$.
Its spectral shape is assumed to be always $\propto \nu^{-1} \exp(-h\nu/150~{\rm keV})$.
The viewing angle $\theta_{\rm v}$ is $3^\circ$ for all states.
The Doppler factor $\delta$ varies from 17.8 (when $\Gamma=13$) to 18.81 ($\Gamma=16$).
{The first four lines show the parameters for the Synchrotron - Inverse Compton blazar model adopted \citep{Ghisellini_Tavecchio2009} to reproduce the radio-to-gamma-ray SEDs of Fig.~\ref{fig:SED_multiepoch_1510}. 
The last line shows the parameters used to reproduce the VHE emission 
of Mar~2009, quasi-simultaneous with lower frequency data (see Fig.~\ref{fig:SED_multiepoch_1510}c).}}
\label{table:modelparams}
\end{table*}



\begin{acknowledgements}
{ We thank the anonymous referee for very helpful comments.}
We are grateful to C. Baldovin, G. Belanger, P. Binko,  M. Cadolle Bel, C. Ferrigno, A. Neronov, C. Ricci,  C. Sanchez, who assisted with the \integral\ observations and quick-look analysis, and to J. Kataoka for sending his 2006 data in digital form.   This work was financially supported by ASI-INAF grant I/009/10/0. This research has made use of data from the MOJAVE database that is maintained by the MOJAVE team \citep{Lister2009,Lister2013}. The MOJAVE Program is supported under NASA Fermi grant NNX12AO87G. This research exploits IRAF package routines \citep{Tody1986,Tody1993}.
{ This research has made use of the NASA/IPAC Extragalactic Database (NED) which is operated by the Jet Propulsion Laboratory, California Institute of Technology, under contract with the National Aeronautics and Space Administration.}
{ EP acknowledges funding from ASI INAF grant I/088/06/0 and from the Italian Ministry of Education and Research and the Scuola Normale Superiore.} AB and MTF thank ASI/INAF for the contract I2013.025-R.0 which financially supported their contribution to the work.
\end{acknowledgements}

\end{document}